\documentclass[superscriptaddress,aps,pra,twocolumn,showpacs,nofootinbib,longbibliography,notitlepage,floatfix]{revtex4-1}
\usepackage{etex}
\usepackage{amsmath,amssymb,amsthm}
\usepackage{hyperref}
\hypersetup{colorlinks=true,citecolor=blue,urlcolor=blue}
\usepackage[pdftex]{graphicx}
\usepackage{times,txfonts}
\usepackage{braket}
\usepackage{color}
\usepackage{amsmath,blkarray}
\usepackage{mathtools}
\usepackage{ulem}
\usepackage{latexsym}
\usepackage{tabularx, booktabs}
\usepackage{graphics,epstopdf}
\usepackage{graphicx}
\usepackage{float}
\usepackage{amsfonts}
\usepackage{tikz}
\usepackage{color,soul}

\newcommand{\be}{\begin{equation}}
\newcommand{\ee}{\end{equation}}
\newcommand{\ba}{\begin{eqnarray}}
\newcommand{\ea}{\end{eqnarray}}

\usepackage{multirow}
\usepackage{appendix}
\usepackage{url}

\begin{document}


\title{Shortcuts to Adiabaticity in Digitized Adiabatic Quantum Computing} 

\author{Narendra N. Hegade}
\affiliation{International Center of Quantum Artificial Intelligence for Science and Technology~(QuArtist) \\ and Physics Department, Shanghai University, 200444 Shanghai, China}

\author{Koushik Paul}%
\email{koushikpal09@gmail.com}
\affiliation{International Center of Quantum Artificial Intelligence for Science and Technology~(QuArtist) \\ and Physics Department, Shanghai University, 200444 Shanghai, China}

\author{Yongcheng Ding}
\affiliation{International Center of Quantum Artificial Intelligence for Science and Technology~(QuArtist) \\ and Physics Department, Shanghai University, 200444 Shanghai, China}
\affiliation{Department of Physical Chemistry, University of the Basque Country UPV/EHU, Apartado 644, 48080 Bilbao, Spain}

\author{Mikel Sanz}
\affiliation{Department of Physical Chemistry, University of the Basque Country UPV/EHU, Apartado 644, 48080 Bilbao, Spain}
\affiliation{IKERBASQUE, Basque Foundation for Science, Plaza Euskadi 5, 48009 Bilbao, Spain}

\author{F. Albarr\'an-Arriagada}
\affiliation{International Center of Quantum Artificial Intelligence for Science and Technology~(QuArtist) \\ and Physics Department, Shanghai University, 200444 Shanghai, China}

\author{Enrique Solano}
\affiliation{International Center of Quantum Artificial Intelligence for Science and Technology~(QuArtist) \\ and Physics Department, Shanghai University, 200444 Shanghai, China}
\affiliation{Department of Physical Chemistry, University of the Basque Country UPV/EHU, Apartado 644, 48080 Bilbao, Spain}
\affiliation{IKERBASQUE, Basque Foundation for Science, Plaza Euskadi 5, 48009 Bilbao, Spain}
\affiliation{IQM, Nymphenburgerstr. 86, 80636 Munich, Germany}

\author{Xi Chen}
\email{xchen@shu.edu.cn}
\affiliation{International Center of Quantum Artificial Intelligence for Science and Technology~(QuArtist) \\ and Physics Department, Shanghai University, 200444 Shanghai, China}
\affiliation{Department of Physical Chemistry, University of the Basque Country UPV/EHU, Apartado 644, 48080 Bilbao, Spain}

\date{\today}

\begin{abstract}

    Shortcuts to adiabaticity are well-known methods for controlling the quantum dynamics beyond the adiabatic criteria, where counter-diabatic (CD) driving provides a promising means to speed up quantum many-body systems. In this work, we show the applicability of CD driving to enhance the digitized adiabatic quantum computing paradigm in terms of fidelity and total simulation time. We study the state evolution of an Ising spin chain using the digitized version of the standard CD driving and its variants derived from the variational approach. We apply this technique in the preparation of Bell and Greenberger-Horne-Zeilinger states with high fidelity using a very shallow quantum circuit. We implement this proposal in the IBM quantum computer, proving its usefulness for the speed up of adiabatic quantum computing in noisy intermediate-scale quantum devices.
    
\end{abstract}

\maketitle

\section{\label{intro} introduction}
Quantum computing is known to have significant advantages in solving certain computational tasks, such as simulating quantum systems \cite{feynman1999simulating, georgescu2014quantum, houck2012chip, lanyon2011universal, gerritsma2010quantum}, machine learning \cite{biamonte2017quantum, lloyd2013quantum, dunjko2016quantum, schuld2019quantum}, solving optimization problems \cite{moll2018quantum, farhi2016quantum, arute2020quantum}, cryptography \cite{gisin2002quantum, BENNETT20147}, and several others. Recent advancements in quantum technologies have already shown that quantum computers can outperform currently existing classical computers \cite{arute2019quantum}. 

Quantum adiabatic algorithms (QAA) \cite{farhi2001quantum, ambainis2004elementary, young2010first, peng2008quantum} are one of the leading candidates for solving optimization problems \cite{ reichardt2004quantum, steffen2003experimental, bapst2013quantum}. In adiabatic quantum computation (AdQC), we start with a simple Hamiltonian whose ground state can be easily prepared and evolve the system adiabatically to the ground state of the final Hamiltonian, which encodes the solution of the optimization problem. This is embodied by the well-known method of quantum annealing \cite{das2008colloquium}. Quantum annealers, such as the D-Wave machine \cite{johnson2011quantum}, provide the test-bed for adiabatic algorithms \cite{boixo2013experimental}. Despite its applications, quantum annealers have certain limitations, such as difficulty in implementing non-stoquastic Hamiltonian, limited qubit connectivity and noise. Although AdQC is equivalent to the standard circuit model \cite{aharonov2008adiabatic}, the advantage of digital quantum computation (DQC) over quantum annealers is that the circuit model offers more flexibility to construct arbitrary interactions, and it is consistent with error correction. The recent work by R. Barends et al. \cite{barends2016digitized} combines the advantage of AdQC and the circuit model, termed as digitized adiabatic quantum computation (DAdQC), has been presented and implemented on a superconducting system.

QAAs are generally governed by the quantum adiabatic theorem that restricts a system to evolve along a specific eigenstate, i.e., from the ground state of an initial Hamiltonian $\hat{H}_i$ to the ground state of a final Hamiltonian $\hat{H}_f$, while the evolution is considerably slow. The computation time for the QAA depends on the minimum energy gap between the successive eigenstates during the evolution. When the system size increases, this poses a significant  disadvantage for the implementation of QAA as the energy gap decreases with increasing system size, which ends up in transition between various instantaneous eigenstates. One has to increase the adiabatic evolution time to circumvent such an issue. However, in practice, evolution time for QAA is significantly larger than the coherence time of the current quantum computers, leading to the loss of fidelity of the evolution.  

The techniques of ``Shortcut to adiabaticity" (STA) \cite{STArev1,STArev2} have been developed during the past decade and proved to be extremely useful for accelerating quantum adiabatic processes in general \cite{takahashi2019hamiltonian}. Various techniques like counter diabatic (CD) driving (equivalently  transitionless quantum algorithm) \cite{demirplack1,demirplack2,berry}, invariant based inverse engineering \cite{ChenPRL104,mugaLRI}, fast-forward approach \cite{FF1,FF2} are rigorously explored and implemented in several studies \cite{SchaffPRA2010,an2016shortcuts,du2016experimental}. Among these works, studies related to quantum spin systems such as Ising and Heisenberg spin models are of particular interest due to their relevance to the applicability and ease of implementation in the development of modern-day quantum algorithms \cite{lucas2014ising}. In particular, the CD driving has been useful for studying fast dynamics \cite{del2012assisted, Steering, takahashi2013transitionless, hartmann2019rapid, petiziol2018fast}, preparation of entangled states \cite{opatrny2014partial, petiziol2019accelerating,ji2019experimental,zhou2020experimental}, quantum annealing \cite{vinci2017,takahashi2017shortcuts,FazioPRR2020} and, others. 

In this work, we explore the STA techniques to enhance the performance of the DAdQC using Ising spin systems. Starting from a single spin, we extend our study to many spins with nearest-neighbor interactions using the CD driving. We find out that the CD interactions in the QAA improve the fidelity remarkably compared to the previous studies. Due to the difficulty in the implementation of the exact CD term for a many-body system, we opt to find local CD terms that can drive smaller systems precisely and able to achieve the target state in very few time steps. By considering approximate CD terms using the nested commutator, we study the non-integrable Ising-model and extend this idea for the preparation of Bell state and GHZ state in larger systems.  

The paper is organized as follows. In the next Sec.~\ref{oneQ}, we give a detailed insight into the implementation of STA in a single spin, where the CD interaction can be exactly calculated using Berry's formula. Sec.~\ref{twoQ} explains the application of the approximate local CD driving and its limitations when applied to strongly interacting many-spin systems. In Sec.~\ref{variTion}, we show the improvement in fidelity when the approximate CD driving is calculated using the nested commutator through the adiabatic gauge potential using the variational approach. This is followed by an example of entangled state preparation, where we show the preparation of Bell state and GHZ state for few spin systems with high fidelity. In the final Sec.~\ref{con}, we summarize our findings and discuss the scope for future research.
\section{\label{oneQ} Single spin system}
We begin our heuristic discussions with a single spin in the presence of  time-dependent external magnetic-field $\boldsymbol{h}(t)$, represented by a two-level Hamiltonian, given by
\begin{equation}
    \hat{H}^{(1)}_0(t) = \boldsymbol{h}(t) \cdot \boldsymbol{\hat{\sigma}},
    \label{1q_ham}
\end{equation}
where $\boldsymbol{\hat{\sigma}}$ represents the Pauli matrices, and the superscript $1$ represents the number of spin. Following the general method for AdQC, also that of quantum annealing, we express this Hamiltonian as a combination of two time-independent parts.  
\begin{equation}
\hat{H}^{(1)}_0\left(t \right)=(1-\lambda(t)) \hat{H}_{i} + \lambda(t) \hat{H}_{f},
\label{Ham_sup}
\end{equation}
$\hat{H}_i$ and $\hat{H}_f $ are time-independent with ground states $\ket{\psi_i}$ and $\ket{\psi_f}$, respectively. The time dependence of the system is introduced through the parameter $\lambda(t)$. The initial Hamiltonian is chosen as $\hat{H_i}=h_x \sigma_{x}$, and the final Hamiltonian as $\hat{H_f}=h_z \sigma_{z}$, where $h_x$ and $h_z$ being the magnetic field strength along respective directions. Such a choice leads to express the magnetic field effectively as $\pmb{h}(t) = [h_x (1-\lambda(t)) \quad 0 \quad h_z \lambda(t)]^T$. AdQC, in its rudimentary approach, allows $\lambda(t)$ to be any function that varies from $0$ to $1$ and drives the system from $\ket{\psi_i}$ to $\ket{\psi_f}$. Although, the most general way to choose it as a linear function, here to begin with, it is considered as $\lambda(t) = \sin^2{(\omega t)}$, where $\omega = \pi / 2 T$ with $T$ being the total evolution time. Although $\hat{H}_0(t)$ is extremely elementary and can easily be implemented in the circuit model \cite{sieberer2019digital}, there are hints that the evolution can be improved significantly using the STA \cite{Steering}. In this case, one should be tempted to find out the CD term, which is somewhat straightforward to calculate using \cite{berry}, 
\begin{figure}
    \includegraphics[width=\linewidth]{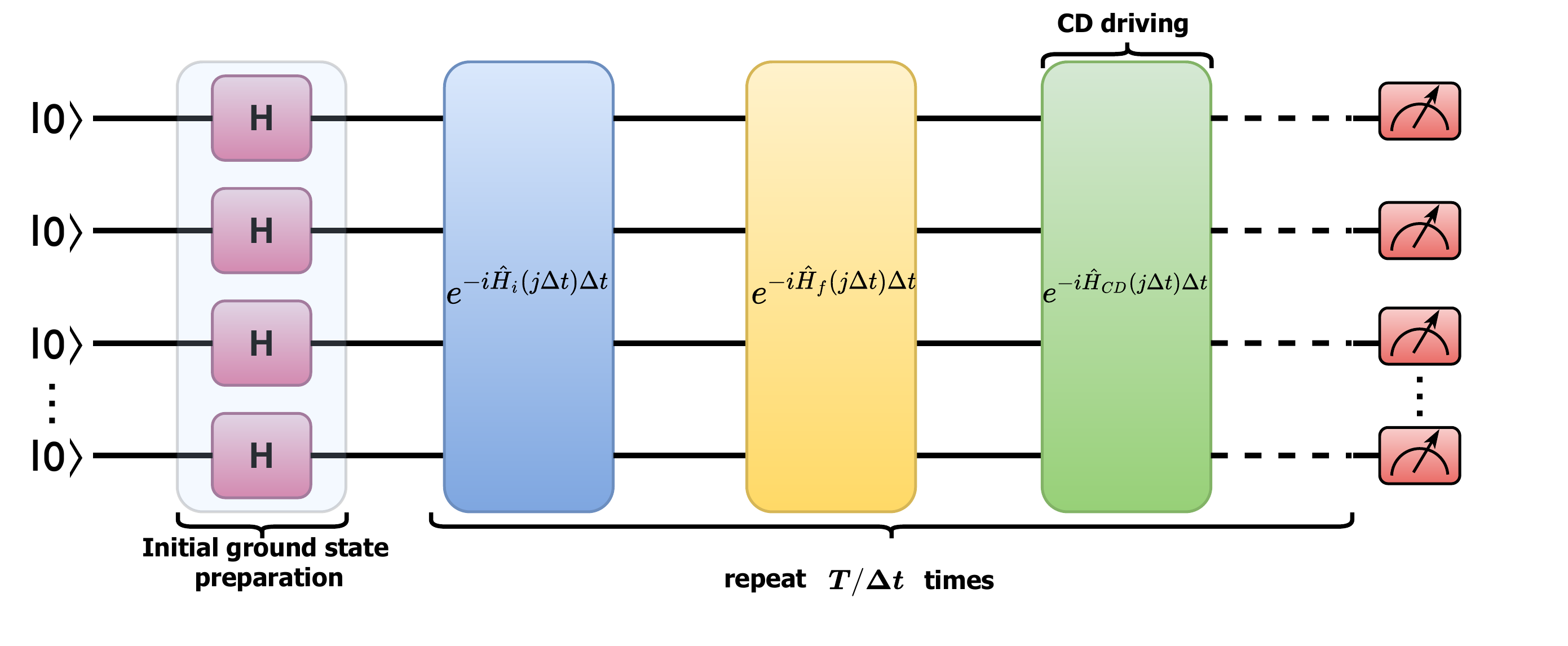}
    \caption{Circuit implementation for the digitized adiabatic evolution using CD driving, where $T$ is the total evolution time, and $\Delta t$ is the step size. The circuit is repeated $n=T/\Delta t$ times, where the Hamiltonian's satisfy the condition $\hat{H}_f(0)= \hat{H}_{CD}(0)= \hat{H}_{CD}(T)=\hat{H}_i(T)=0$.}
    \label{fig1}
\end{figure}
\begin{equation}
    \hat{H}^{1}_{cd}(t) = \frac{1}{2 |\boldsymbol{h}(t)|^2} (\boldsymbol{h}(t) \times \dot{\boldsymbol{h}}(t))\cdot\boldsymbol{\hat{\sigma}},
    \label{CD_gen}
\end{equation}
yielding the explicit form as following,
\begin{align}
\hat{H}^{(1)}_{cd}(t)= F^{(1)}(t) \sigma_{y}= \frac{-h_x h_z \partial_t ( 1- \lambda(t))}{2\left[h_x^{2} (1-\lambda(t))^2 + h_z^{2} \lambda^2(t)\right]} \sigma_{y}.
\label{CD_spec2}
\end{align}
Therefore, the total Hamiltonian assisted with CD term becomes 
\begin{equation}
 \hat{H}^{(1)}(t) = (1-\lambda(t)) h_x \sigma_{x} + \lambda(t) h_{z} \sigma_{z} + F^{(1)}(t) \sigma_y.
 \label{Ham_F}
\end{equation}
Note that the introduction of the CD term should not affect the initial and final states, since $F^{1}(t)$ should always satisfy the boundary conditions, $F^{(1)}(t=0) = F^{(1)}(t=T) = 0$. Also, the STA methods generally follow the inverse engineering approach of quantum control, i.e., designing the interaction for achieving the desired eigenstates. Therefore the notion of the eigenstate, although not that essential in traditional AdQC, turns out to be extremely important in the present case. Here, the initial state of $\hat{H}_i$ is chosen in the computational basis, i.e., $\{ \ket{0},\ket{1}\}$ , as $\ket{\psi_i} = \ket{+}$ and the final state is, $\ket{\psi_f} = \ket{1}$. It should be noted that $\ket{+}=(\ket{0}+\ket{1})/\sqrt{2}$ and $\ket{1}$ are the natural ground states of $\hat{H}_i$ and $\hat{H}_f$, respectively, but these choices are not restricted to the ground states only. However, such a choice restraints the qubit in the ground state, minimizing the effect of the decoherences. 

\begin{figure}
    \begin{center}
    \includegraphics[width=\linewidth]{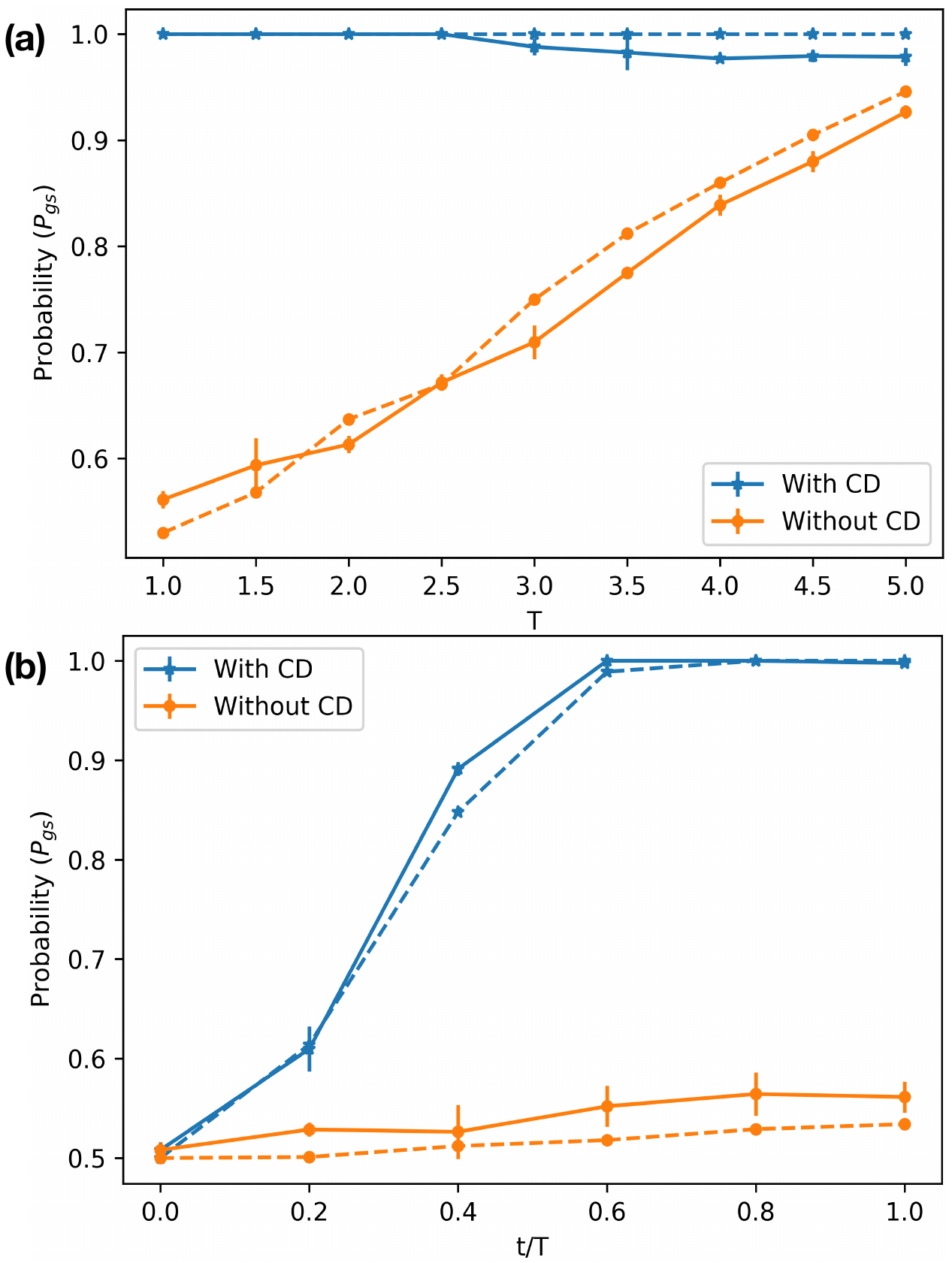} 
    \caption{(a) The final ground state probability $P_{gs}$ versus the simulation time for a single qubit using CD driving on \texttt{ibmq\_essex} quantum computer (solid blue) compared to the ideal simulator (dashed blue). The simulation without CD driving in real device (solid brown) and ideal simulator (dotted brown). (b) Time evolution using DAdQC and STA methods for $T=1$. Parameters are following: $\Delta t = 0.2$, $h_x =-1$, $h_z = 1$, and number of shots $(N_{shots})=1024$.}
    \label{fig2}
    \end{center}
\end{figure}
To implement the evolution using one qubit, we used first-order Trotter-Suzuki formula. The time evolution is digitized with $n$ number of small time steps $\Delta t$ (see Eq~\eqref{trotter}). Ideally, the discretized version of AdQC approaches the actual adiabatic evolution for $n=T/\Delta t \to \infty $, ($\Delta t \to 0$).
Although, in real situations, $n$ is finite and it has to be a relatively small number since each trotter step is being implemented by three rotation gates (see Appendix \ref{app1}). The error associated with the first-order trotterization is $\mathcal{O}({\Delta t}^2)$ \cite{suzuki1976generalized}.

To perform the simulation, we used publicly available five qubit superconducting quantum computer of IBM Quantum Experience \cite{ibmq}. For the single spin experiment we use qubit $Q_0$ on $ibmq\_essex$. Since the single qubit gate error of the device is of the order of $10^{-4}$, the initial state is prepared with very high fidelity. 
Also a significant error in this simulation comes from the readout error $(\sim4\%)$, for that we used the error mitigation technique using matrix inversion method described in Appendix \ref{app3}.

In the simulation, time evolution takes place between $\ket{+}$ to $\ket{1}$ and measured in the computational basis, which restricts the variation of the probability between $0.5$ to $1$, see Fig.~\ref{fig2}. Since the trotter error is of the order $\mathcal{O}(\Delta t ^2)$, we choose, $\Delta t=0.2$, and $T=1$ for the comparison of the evolution using DAdQC and STA assisted DAdQC. To study the probability of final ground state $P_{gs} = |\braket{\psi(t)|1}|^2$, where $\ket{\psi(t)}$ denotes the time evolved state of $\hat{H}_0(t)$, measurement is performed at each progressing time step. Fig.~\ref{fig2}(b) shows a single evolution of the qubit governed by both Eq.~\eqref{Ham_sup} and Eq.~\eqref{Ham_F}, using both simulation and experiment on a real quantum device. With the application of CD driving $F^{(1)}(t)$, the probability of getting the state $\ket{1}$ from the experiment comes out to be around $0.997$. Whereas, when the CD driving is zero, i.e., for the adiabatic evolution, the final state probability is around $0.561$ only. However, if the evolution is extended for a larger $T$, we could obtain a much higher probability, even without $F^{(1)}(t)$, a signature of the typical adiabatic process. This is evident in Fig.~\ref{fig2}(a), where the fidelity of the evolution (in the computational basis) for different $T$ is shown. Even when $T=1$ ($\Delta t = 0.02$) using the STA method, the final ground state is reached with nearly unit fidelity. 
We observe that fidelity for the STA method for large $T$ maintains its value at around $0.978$. However, for the adiabatic case, fidelity gradually increases with increasing simulation time and the average fidelity will be around $0.927$ for $T=5$. Notice that in Fig.~\ref{fig2}(a), the experimental values differ slightly from the exact simulation values, and the difference is slightly larger for the STA assisted case. As $T$ increases, the circuit depth becomes larger, which results in ramping up the gate errors, affecting STA more than the adiabatic case as it requires more gates for implementation.
\section{LOCAL COUNTER-DIABATIC DRIVING} \label{twoQ}
The results in Sec.~\ref{oneQ} establishes the fact that STA assisted DAdQC shows significant improvement over the DAdQC, at least when a single qubit is considered. However, such implementation becomes far more interesting when multiple qubits are considered. 
The simplest choice is a system of $N$ interacting spins in one-dimensional lattice, coupled by a time-dependent exchange interaction $J(t)$ with a rotating magnetic field is acting upon it. Here we consider $J(t)$ to constitute $\sigma_z\sigma_z$ type interaction with $J_0$ being the coupling amplitude. The spins are initially aligned along the transverse magnetic field, $h_x$, while an Ising Hamiltonian represents the system's final state. The total Hamiltonian is represented as, 
\begin{equation}
    \hat{H}^{(N)}_0(t) = (1-\lambda(t)) \sum_{j=1}^{N} h_x \sigma_{x}^j
     + \lambda(t)\sum_{j=1}^{N} ( h^j_z  \sigma_{z}^j  + J_0 \sigma_{z}^j \sigma_{z}^{j+1}).
     \label{non-int}
\end{equation}
The scheduling $\lambda(t)$ is chosen similarly, as in Eq.~\eqref{Ham_sup}. The traditional approaches to finding the CD driving are predominantly limited to two and three-level systems and become more complex for higher dimensional many-body systems. However, for interacting many spin systems, as in the preceding section, a local CD driving could be more useful. Instead of acting on the whole system, a set of approximated interactions could be designed to control the spins individually. Such type of local CD driving is more general and can be extended to a larger number of spins. 
\subsubsection{Local CD from Berry's algorithm}
To realize such CD driving, it is intuitive to approximate the system to a non-interacting one. Using mean-field approximation, this can be achieved effectively for an infinite-range Ising model \cite{hatomura2017shortcuts}. However, this is problematic for DAdQC, as it requires self-consistent feedback $\braket{\sigma_j}$ after every step. Instead, we consider a more direct approach. Since, at $t = 0$, the spins have no mutual interaction and are dictated by the transverse magnetic field, it can be assumed that, during the evolution, the magnitude of $h^j_z$ and $J_0$ grows gradually from zero to some maximum value while the system evolves gradually from $\hat{H}_i$ to $\hat{H}_f$. Therefore, we approximate that those spins are governed by a local effective magnetic field, 
given by ${h}(t) = [h_x (1-\lambda(t)) \quad 0 \quad \tilde{h}^j_z \lambda(t)]^T$, where $\tilde{h}^j_z = h^j_z + J_0$. Subsequently, the local CD driving is calculated and summed over for each spin using Eq.~(\ref{CD_gen}),
\begin{align}
\hat{H}^{(N)}_{CD}(t)=\sum_{j=1}^{N} F^{(N)}_j(t) \sigma_{y}^{j}=\sum_{j=1}^{N} \frac{-h_x \tilde{h}^j_z \partial_t ( 1- \lambda(t))}{2\left[h_x^{2} (1-\lambda(t))^2 + (\tilde{h}^j_z)^{2} \lambda^2(t)\right]} \sigma_{y}^{j}.~~~~~~
\label{CD_spec2q}
\end{align}
Therefore, the modified Hamiltonian that governs the evolution can be expressed as
\begin{equation}
     \hat{H}^{(N)}(t) = \hat{H}^{(N)}_0(t) + \sum_{j=1}^{N} F^{(N)}_j(t) \sigma_{y}^{j}.
     \label{Ham2qTot}
\end{equation}
\begin{figure}
    \centering
    \includegraphics[width=\linewidth]{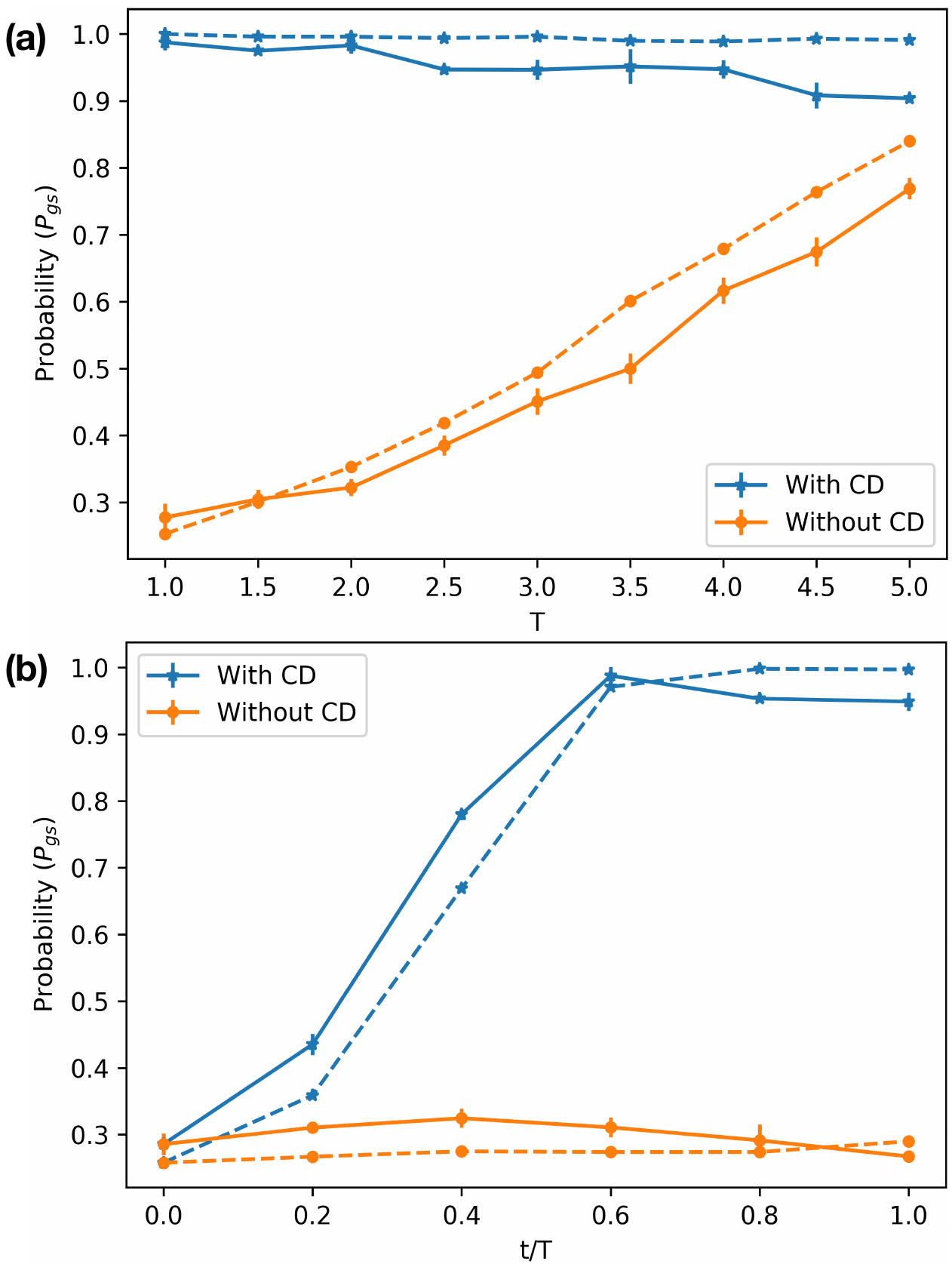}
    \caption{(a) The final ground state probalility $P_{gs}$ versus the simulation time for the two interacting spin system ($\Delta t=0.5$). The red solid curve represents the time evolution using STA method, the solid blue curve represents the DAdQC on \texttt{ibmq\_london} 5-qubit quantum processor. As the evolution time increases the gate error starts to dominate, which is clearly inferred from the figure. (b)Implementation of the time evolution for two spin system without CD term (solid blue) and including CD term (solid red). Parameters are following: simulation time $T = 1$, $\Delta t = 0.2$, $h_x = -1, h_z^1 = h_z^2= 1$, $J_0= -0.1$, and $N_{shots}=1024$. Both the curves are showing expected profile.}
    \label{fig:3}
\end{figure}
As an example, we consider interacting two qubit system, where the time evolution for $\hat{H}^{(2)}(t)$ can be easily implemented by two qubit entangling gates and single qubit rotation gates. The general circuit for implementing the evolution using CD driving is shown in Fig.~\ref{fig1}. The initial and the target states chosen for the evolution are $\ket{++}$ and $\ket{11}$, respectively, which is inferred directly from the following parameters: $h_0=-1$, $h_1=h_2=1$ and $J_0=-0.1$.
The time evolution for STA assisted DAdQC and DAdQC for two qubits are shown in Fig.~\ref{fig:3}. Again, like the single qubit case, one can achieve a high fidelity for the target state preparation. The result obtained from the ideal digital simulator, in Fig.~\ref{fig:3}(a), shows that, when the additional term $\hat{F}^{(2)}_j(t)$ is considered, the target state can be achieved with almost unit fidelity. Furthermore, when a single evolution is considered, as depicted in Fig.~\ref{fig:3}(b), the target state can be achieved substantially faster than adiabatic evolution. However, when implemented in the real experiment, fidelity around $0.93$ is achieved with the application of the CD term, where the fidelity in the computational basis is calculated as $|\braket{\psi_i(n\Delta t)|\psi_f}|^2$. The application of the CD term is more suitable when the evolution time $T$ is small. In principle, the fidelity should remain the same even if we increase the number of time steps. Nevertheless, due to limited coherence time and the increasing number of gates required to implement the CD term, fidelity gradually decreases as depicted in Fig.~\ref{fig:3}(a). 
\subsubsection{Local CD from variational approach}
\begin{figure}
    \centering
    \includegraphics[width=\linewidth]{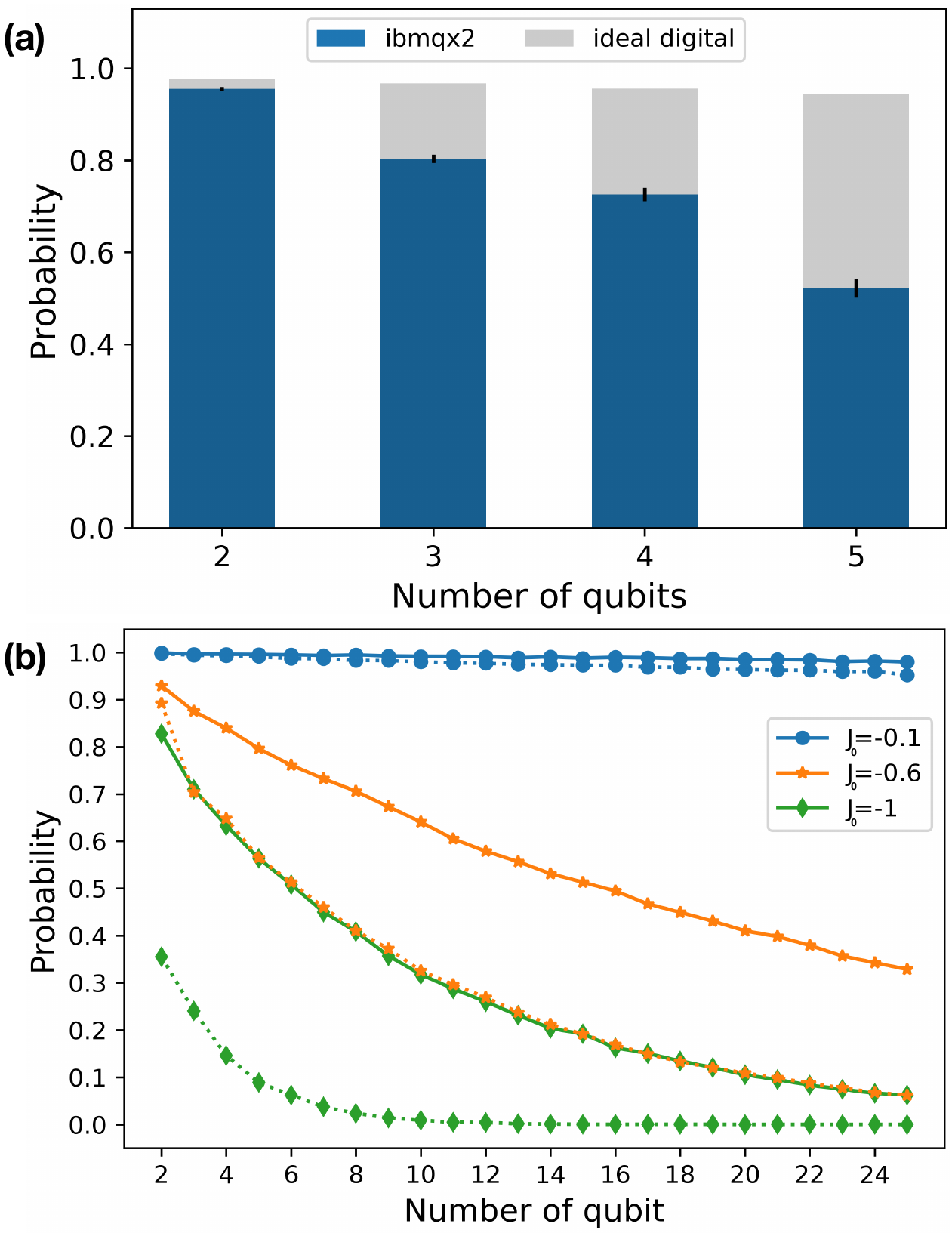}
    \caption{ (a) For the non-integrable Ising model, the probability of obtaining the final ground state using local CD term calculated from variational method for up to 5 qubit system is depicted. The experiment is performed on 5 qubit \texttt{ibmqx2} processor. The experimental parameters are $J_0=-0.1$, $h_x=h_z=1$, $N_{shots}=8192$. (b) The probability of obtaining the final ground state as a function of coupling strength using local CD driving for up to 25 qubits are shown. The solid line is for the local CD term from the variational approach, and the dotted line is for the local CD term from Berry formula, see Eq.~\eqref{CD_spec2q}. The parameters chosen are $h_x=-1$, $h_z^j=1$, $dt =0.1$, and $T=1$. The simulation was performed on a \texttt{qasm\_simulator}.}
    \label{localCD}
\end{figure}

A recently proposed method by Sels and Polkovnikov \cite{sels2017}, based on the variational approach, also provides an alternative to calculate the approximate CD Hamiltonian with only local terms. The method for this calculation is to choose an appropriate adiabatic gauge potential $\hat{A}_\lambda ^* $ \cite{kolodrubetz2017geometry} and minimizing the action
$S=\operatorname{Tr}\left[\hat{G}_{\lambda}^{2}\right] $, where the operator $\hat{G}_\lambda$ is defined by $\hat{G}_\lambda = \partial_\lambda \hat{H} + i [\hat{A}_\lambda ^*,\hat{H}]$ (see Appendix \ref{app2}). The CD driving using this method is expressed as $\hat{H}^{(N)}_{CD} = \dot{\lambda} \hat{A}_{\lambda}^*$. Since the Hamiltonian contains only real values in the $z$-basis, the simplest ansatz is to choose ${\hat{A}}_{\lambda}^{*}=\sum_{j} \alpha_{j}(t) \sigma_{y}^{j}$, i.e., applying an additional magnetic field along the y-direction for each spin. By minimizing the action $S$ with respect to $\alpha_j$, the variational coefficient $\alpha_{j}(t)$ is analytically calculated, which takes the general form, for Eq.~\eqref{non-int}~\cite{sels2017},
\begin{equation}
    \alpha_j(t) = \frac{1}{2}\frac{h_x {h}^j_z}{\left[h_x^{2} (1-\lambda(t))^2 + ({h}^j_z + 2 J_0)^{2} \lambda^2(t)\right]}.
    \label{var_coef}
\end{equation}
The expression for $\hat{H}^{(N)}_{CD}=\sum_{j} \dot{\lambda} \alpha_{j}(t) \sigma_{y}^{j}$ is similar to that of Eq.~\eqref{CD_spec2q} except for a few modifications. In Fig.~\ref{localCD}(a), the probabilities of obtaining the ground state, from both the ideal simulator and the experimentally implemented data from \texttt{ibmqx2} are shown for up to 5 qubits. 
Like the previous case, the final ground state $\ket{11..1}$ can be prepared using the additional $\hat{H}^{(N)}_{CD}$ with high fidelity. The ideal simulator data shows that the final probability, $|\braket{\psi(T)|11..1}|^2$ reaches almost unity for $T=1$ in five trotter steps, especially when $|J_0|$ is small. However, the implemented value differs from the simulator due to the device errors.   

It should be noted that, in the above discussion, the interaction strength $J_0$ is kept sufficiently small compared to the external magnetic field, $J_0 \ll h^j_z$. The ground state of the final Hamiltonian is a ferromagnetic state, i.e., either $\ket{00..0}$ or $\ket{11..1}$, depending on the sign of $h_z^j$. In such scenario, the evolution assisted by the local CD terms in Eq.~\eqref{CD_spec2q} and Eq.~\eqref{var_coef} produces the exact final ground state. Fig.~\ref{localCD}(b) compares the probability of interacting multi-qubit system with ground state $\ket{111\dots1}$. 
For $J_0 \ll h^j_z$, the probability is around $98\%$ in the ideal simulator for both the methods. When $J_0$ becomes higher, the probability using Eq.~\eqref{CD_spec2} decreases drastically and reduces to $35\%$, even for two qubits. Whereas, for the variational approach, the probability is significantly higher for large $|J_0|$ values. When the ground state is degenerate, the obtained result seems to differ from the actual ground state. For instance, when $J_0=2$, with the similar values of other parameters, the ground state becomes doubly degenerate,i.e., the states $\ket{01}$ and $\ket{10}$, the application of the CD term drives the system to one of the eigenstates. This comes out to be true for many spin systems also. Moreover, the calculation of the local CD term is based on the approximation that every spin is treated individually by considering an effective magnetic field acting upon each spin. The effects of the interaction $J_0$ are undermined while calculating the CD term. As a result, the CD term does not help the system evolve into the exact ground state when $J_0$ is comparable or stronger than that of the local magnetic field.  

Subsequently, when $h_z^j = 0$, the final ground state becomes entangled, and one can deduce from Eq.~\eqref{CD_spec2q} that for small $J_0$, the CD term becomes small , i.e, $\hat{F}^{(N)}_j(t)\rightarrow 0$. 
In such cases, the final evolved state, in a short $T$, does not match the adiabatic one. For the variational approach, the CD term vanishes altogether and can not be applied using such form. Therefore, if we are to prepare a highly entangled state, the single qubit approximation for the CD term is not a good choice. 
This drawback occurs as 
CD driving is calculated using the $\sigma_y$ terms, which refers to driving a single qubit with the external magnetic field only. In fact, the spin-spin interaction term decides the final state here and the driving for $\sigma_z \sigma_z$ coupling has to be incorporated. This enforces the fact that the direct approach from the first principle to find the local CD driving is not realistic and should contain other interactions such as $\sigma_y\sigma_z$ and $\sigma_z \sigma_x$ etc. \cite{vinci2017}.
\section{\label{variTion}  Approximate counter-diabatic driving}
Following the discussion in the preceding section, when complex many-body systems are considered, the calculation of the exact CD term becomes difficult. Also, the form of the CD term can be severely complicated with different non-local and many-body interaction terms. Besides, it becomes rather difficult to implement systems with such interactions on current quantum computers. Although, the local terms, see Eq.~\eqref{var_coef}, from variational approach gives an optimal solution, it is not that useful for preparing entangled states, especially when $h_z^{(j)} = 0$. The nature of the CD term from variational calculation depends on the choice of appropriate adiabatic gauge potential $\hat{A}_\lambda ^*$. A recently proposed method gives a more general way to choose the gauge potential by using the nested commutator (NC)~\cite{sels2019prl}, 
\begin{equation}
    \hat{A}_{\lambda}^{(l)} = i \sum_{k = 1}^l \alpha_k(t) \underbrace{[\hat{H},[\hat{H},......[\hat{H},}_{2k-1}\partial_{\lambda} \hat{H}]]],
    \label{gauge}
\end{equation}
where $l$ determines the order of the expansion. Depending on the required accuracy, we can keep the number of variational coefficients small. If we consider only the first-order term, our ansatz will be  $\hat{A}_{\lambda}^{(1)} = i \alpha_1(t) [\hat{H},\partial_{\lambda} \hat{H}]$, and the effective Hamiltonian can be written as, 
\begin{equation}
    {\hat{H}}_{eff}(t)={\hat{H}}(\lambda)+\dot{\lambda} {\hat{A}}_{\lambda}^{(1)},
\end{equation}
where $\dot{\lambda} {\hat{A}}_{\lambda}^{(1)}$ is the relevant CD term.
\begin{figure}
    \centering
    \includegraphics[width =\linewidth]{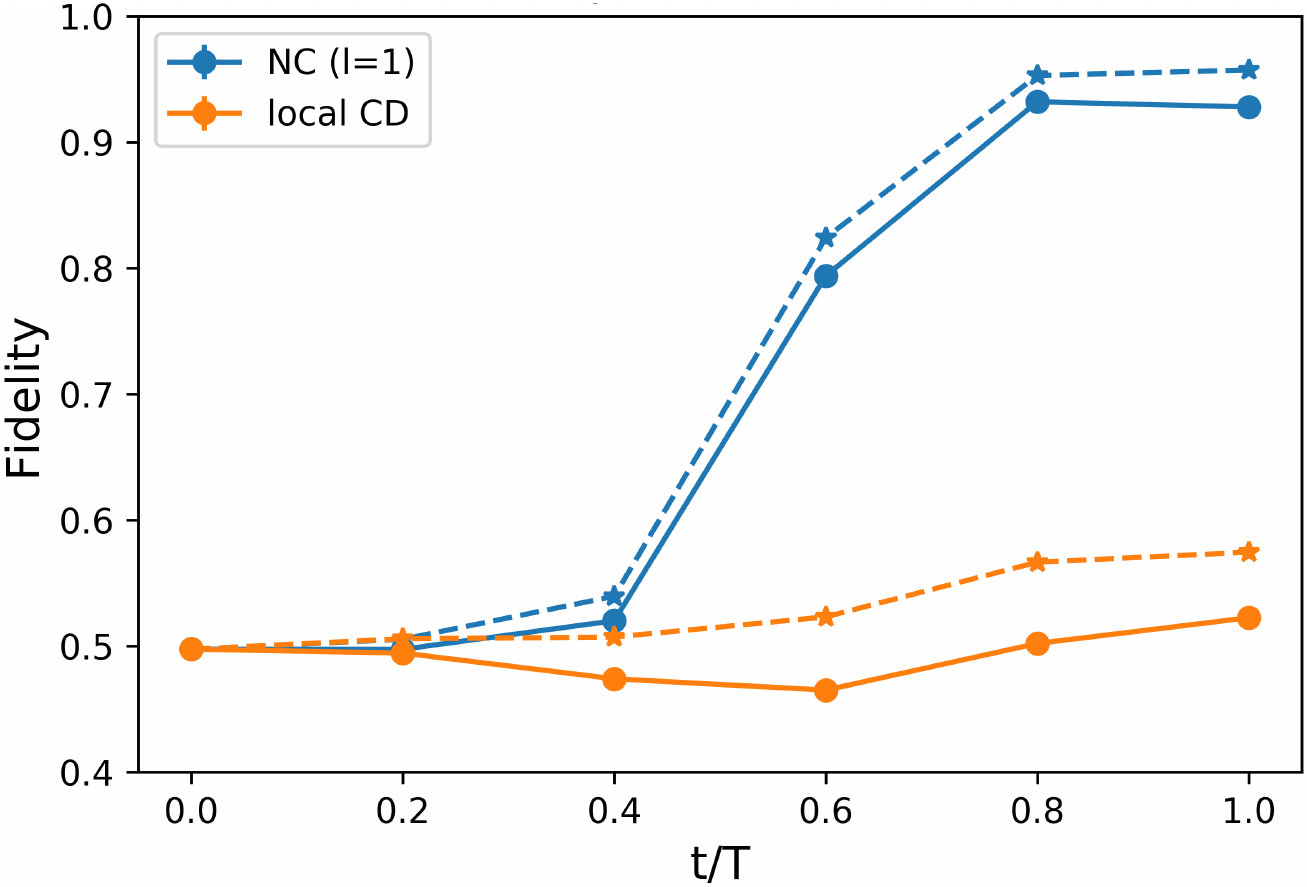}
    \caption{The fidelity of obtaining the final ground state $(\ket{01}+\ket{10})/\sqrt{2}$ as a function of evolution time for local CD and NC ansatz (l=1) obtained from \texttt{ibmq\_vigo}. The solid line represents the experimental result and the dashed line represents the result from ideal digiatal simulator. Parameters: $J_0=2$, $h_z=0.6$, $T=1$, $dt=0.2$, and $N_{shots}=8192$.}
    \label{nc_localCD}
\end{figure}
First of all, we apply this technique to non-integrable Ising spin model, described by the Hamiltonian in Eq. \eqref{non-int}. Considering the two-qubit system ($N=2$), we approximate CD term using first-order nested commutator,
\begin{equation}
    \hat{H}^{(2)}_{CD} = 2\alpha_1(t) h_x \left[ h_z (\sigma_y^1 + \sigma_y^2) + J_0 (\sigma_y^1 \sigma_z^2 + \sigma_z^1 \sigma_y^2 \right],
\end{equation}
with,
\begin{equation}
\alpha_1(t) = \frac{1}{4} \frac {h_z^2 + J_0^2}{ \lambda^2 (h_z^4 J_0^4+ 3h_z^2 J_0^2) +  (1-\lambda)^2 h_x^2 (h_z^2 + 4J_0^2)}.
\label{alpha1}
\end{equation}
The second-order term ($l=2$) can give the exact gauge potential \cite{sels2019prl}. However, for the experimental demonstration, we only consider the first-order term and implement the time evolution on a quantum processor. 
The circuit implementation for the CD driving is shown in Appendix \ref{app1}. Using this method, the final ground state is achieved with very few trotter steps compared to digitized adiabatic evolution, which drastically reduces the number of gates required as well as the total simulation time. In Fig.~\ref{nc_localCD}, we depicted the fidelity as a function of evolution time using first-order nested commutator method when the final ground state is degenerate and compared the result with the local CD term from Eq. \eqref{var_coef}. The fidelity is much better compared to the local CD case for degenerate state and thereby it justifies our argument in the preceding section.

Secondly, we shall check the reliability and validity as well as the extent of the variational approach in the many body regime. To this aim, we apply this technique to prepare the GHZ state in Ising spin chain with many spins, described by the Hamiltonian, 
\begin{equation}
     \hat{H}(\lambda(t)) = (1-\lambda(t)) \sum_j^N h_x \sigma_{x}^j + \lambda(t) J_0 \sum_j^N \sigma_{z}^j \sigma_{z}^{j+1},
     \label{lambdaHam}
\end{equation}
with $N$ being the number of spins. Here, the periodic boundary condition $\sigma^{N+1} = \sigma^0$ is assumed. Following the procedure described in Sec.~\ref{variTion}, by considering only the first-order expansion, we calculate the approximate gauge potential as,
\begin{equation}
    {\hat{A}}_{\lambda}^{(1)}=2 \alpha_{1}^N (t) J_0 h_{x} \sum_j^N \left(\sigma_{z}^{j} \sigma_{y}^{j+1}+\sigma_{y}^{j} \sigma_{z}^{j+1}\right).
\end{equation}
The variational coefficient, $\alpha^N_1(t)$ is calculated by minimizing the action $S$. For the experimental demonstration on a quantum processor, we choose a small system with two and three qubits to prepare a Bell state and GHZ state. For the bell state, $(\ket{00} + \ket{11})/\sqrt{2}$, governed by the Hamiltonian in Eq.~\eqref{lambdaHam}, the variational coefficient is calculated as $\alpha_1 (t) =- J_0 h_{x}/2[J_0^{2}\lambda^2+4(1-\lambda)^{2} h_{x}^{2}]$. Here we have noticed that, for two spins, the first order commutator is proportional to the higher-order terms. The resulting CD driving from the approximate gauge is exact and produces unit fidelity in ideal situations \cite{sels2019prl}. The same procedure can be followed in case of more qubits to prepare a GHZ state $\ket{GHZ}=(\ket{0}^{\otimes N}+\ket{1}^{\otimes N})/
\sqrt{2}$ starting from the $N$ qubit ground state $\ket{+}^{\otimes N}$. Specifically, the variational coefficient for a three-qubit case is given by ${\alpha}_{1}(t)= -J_0 h_{x}/[5 J_0^{2}\lambda^2+8(1-\lambda)^{2} h_{x}^{2}]$.
\begin{figure}
    \centering
    \includegraphics[width=\linewidth]{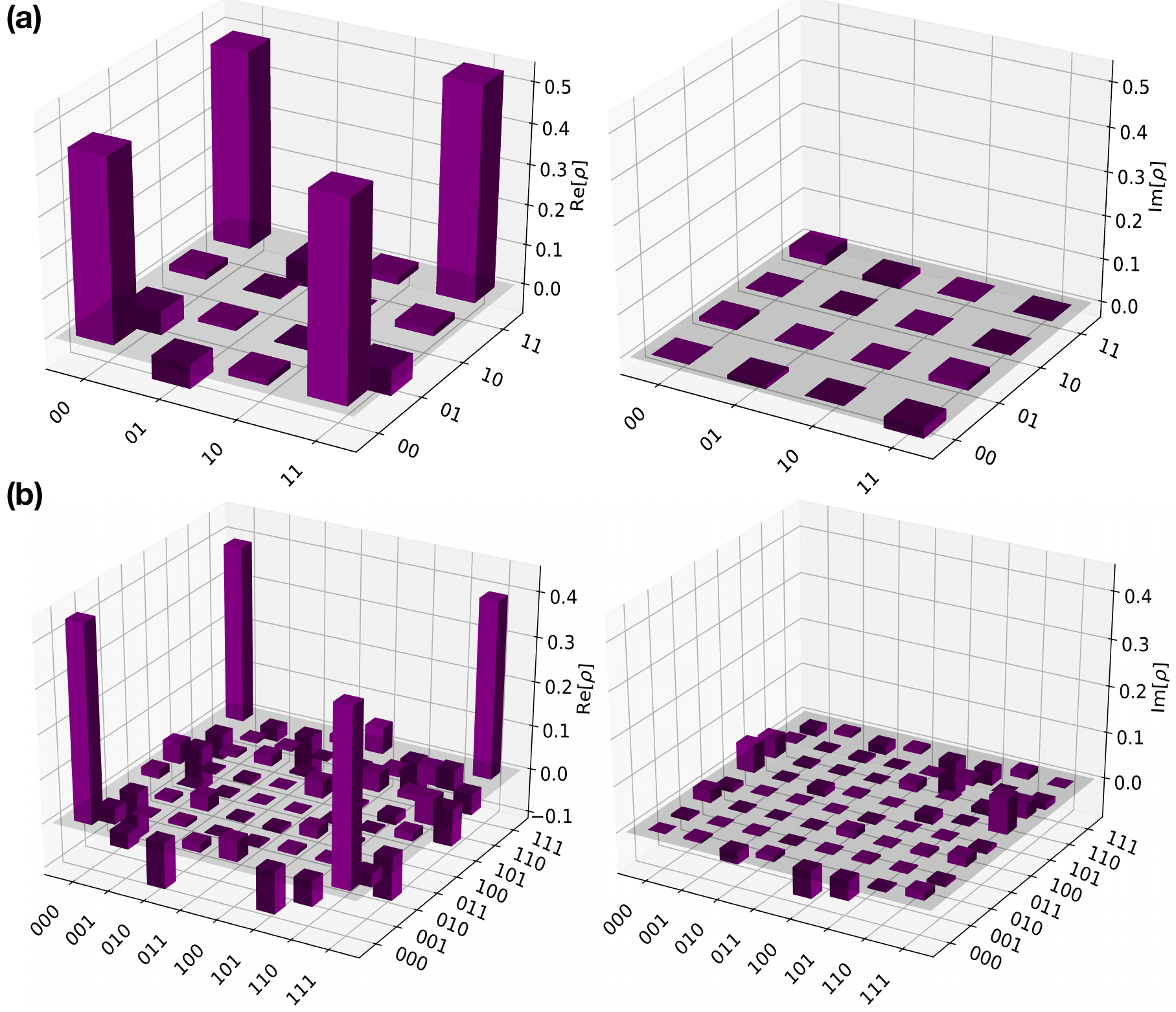}
    \caption{The density matrix representation of the final ground state obtained from state tomography. (a) Bell-state from \texttt{ibmq\_ourense} and (b) GHZ-state from \texttt{ibmq\_vigo}.}
    \label{fig:10}
\end{figure}

\begin{figure}%
    \centering
    \includegraphics[width=\linewidth]{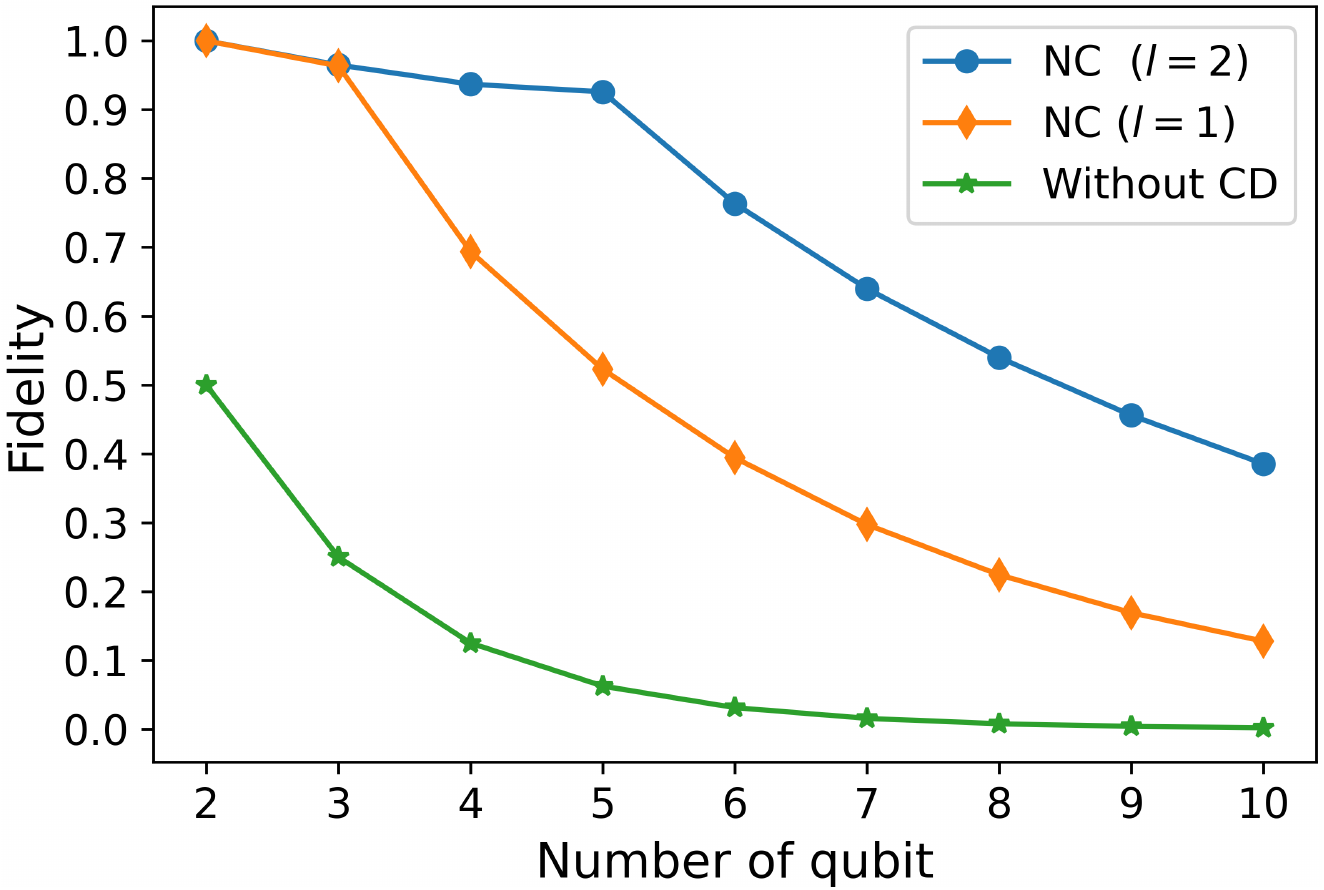}
    \caption{Fidelity to prepare the GHZ state as a function of system size on an ideal digital simulator with CD term from nested commutator (NC) ansatz with different orders and the naive approach without CD term. Where the parameters are $T=0.006$ and $\Delta t=0.001$.}%
    \label{fig:12}%
\end{figure}

The simulation was performed on a five qubit quantum processor \texttt{ibmq\_ourense}. A similar trotterization, as in Eq.~(\ref{trot_final}), is used to study the evolution with digitized time step $dt = 0.01$. Using the CD driving, with only three trotter steps, the desired bell state is obtained with experimental fidelity $0.984$. The ideal digital simulation gives almost unit fidelity ($F=0.999$). The fidelity is calculated as, $F\left(\rho_{1}, \rho_{2}\right) = \braket{\psi_{1}\left|\rho_{2}\right| \psi_{1}}$, where the exact bell state is represented by, $\rho_{1}=\left|\psi_{1}\right\rangle\left\langle\psi_{1}\right|$.  
Similarly, for the three-qubit system, the ideal digital evolution gives the fidelity $0.935$ with the exact GHZ state, and the corresponding experimental fidelity is $0.819$. The density matrix representation of the final state ($\rho_2$) is obtained by performing quantum state tomography for both Bell and GHZ states and is depicted in Fig. \ref{fig:10}. Whereas, Fig.~\ref{fig:12}, shows how the fidelity varies with increasing system size on a ideal digital simulator with six trotter steps. The first-order approximation of the CD term provides high fidelity for small system size. As we increase $N$, the probability of obtaining the final ground state decreases gradually. This can be overcome by considering the higher-order commutators while calculating the approximate CD term. As the variational method tends to provide exact CD driving for larger $l$-values, which in principle, can give a better fidelity in many-body systems.   
\section{\label{con}Conclusion}
In conclusion, we have demonstrated the implementation of digitalized STA on a superconducting quantum processor. The problem Hamiltonian, chosen for the simulation, emulates the one-dimensional Ising spin chain. The STA is realized by means of the local and approximate CD driving, which is obtained using mainly two methods: the long-established Berry's algorithm and the newly proposed variational approach. The CD term in our simulation is non-stoquastic in nature, therefore it can't be simulated efficiently on a classical computer. The effective Hamiltonian is implemented using the available quantum gates, and the time evolution of the system is studied to achieve the ground state of the problem Hamiltonian. We have showed that the time steps required to reach the target state are minimal compared to the DAdQC method, leading to minimal loss due to the decoherences and accumulated gate errors. The local CD driving proved to be very effective for weakly interacting spin chains; however, is not useful in case of strong interaction and fails to produce degenerate target states. To remedy this situation, the approximate CD driving is useful, which can be calculated using the nested commutator method. We applied the first-order approximate CD term to prepare Bell and GHZ state with high fidelity for few qubit systems. Furthermore, for the many qubit systems, the fidelity can be further improved with higher-order terms at the cost of gate numbers.

This work provides the evidence that significant enhancement of the DAdQC approach can be achieved using the STA methods by decreasing the total computational cost and hence achieving the desired results within the coherence time of the device. To our knowledge, this work is the first to successfully realize the STA methods in a contemporary superconducting circuit-based quantum computer. Due to the decoherence and gate errors, the time evolution studies are challenging to perform in such devices. Our result provides a beginning to such studies on the speed up of AdQC algorithms.
\section*{Acknowledgement}
The authors are grateful to Kazutaka Takahashi for useful discussions. We acknowledge support from National Natural Science Foundation of China (NSFC) (11474193), STCSM (2019SHZDZX01-ZX04, 18010500400 and 18ZR1415500), Program for Eastern Scholar, Ram\'on y Cajal program of the Spanish MCIU (RYC-2017-22482), QMiCS (820505) and OpenSuperQ (820363) of the EU Flagship on Quantum Technologies, Spanish Government PGC2018-095113-B-I00 (MCIU/AEI/FEDER, UE), Basque Government IT986-16, as well as the and EU FET Open Grant Quromorphic. This work is also partially supported from Huawei HiQ funding for developing QAOA\&STA (Grant No. YBN2019115204). 
\begin{appendices}
\appendix
\section{Method of digitization}
\label{app1}
The circuit model can efficiently simulate the adiabatic quantum computing by using the digitization of continuous adiabatic evolution. The time-dependent Hamiltonians, considered in this work, can be represented as a sum of $M$ $k$-local terms that act on at most $k$-qubits. This can be represented as
\begin{equation}
    \hat{H}(t)=\sum_{m=1}^{M} C_m(t) \hat{H}_m.
    \label{hm}
\end{equation}
The continuous time evolution operator of $\hat{H}(t)$ is given by, 
\begin{equation}
 \hat{U}\left(0, T\right)=\operatorname{\mathcal{T} exp}\left[-i \int_{0}^T d t \hat{H}(t)\right],  
 \label{Ucont}
\end{equation}
where $\mathcal{T}$ is the time ordering operator. The discretization is done using the first order trotter-suzuki formula,
\begin{equation}
\hat{U}(0, T) \to \hat{U}(0,T)_{dig} = \prod_{j=1}^{n} \prod_{m=1}^M \exp \left\{-i \Delta t C_m(j \Delta t)  \hat{H}_{m}\right\}.
\label{trotter}
\end{equation}
Here the total evolution time $T$  is divided into $n$ equal steps of width $\Delta t$ i.e., $n=T/\Delta t$. In this case, the error would be of the order $\mathcal{O}(\Delta t^2)$~\cite{suzuki1976generalized}. One can also consider higher-order decomposition, which can give better approximation by minimizing the error further \cite{hatano2005finding}. However, an interesting observation from our simulation is that the digital adiabatic evolution using CD driving is independent of the simulation time $T$, and depends only on the number of trotter steps. So, by fixing the total time steps, we can choose an arbitrarily small value for $T$ and $\Delta t$ so that we can achieve arbitrary precision even with the first order trotterization. We ignore the time variation of the Hamiltonian $\hat{H}(t)$ on time scale lower than $\Delta t$, which contributes to an extra error $\sim \|\partial \hat{H} / \partial t\| \Delta t$ per each step. When the Hamiltonian fluctuation is very fast, it is possible to suppress this additional error, which is discussed in \cite{poulin2011quantum}. From Eq.~\eqref{trotter}, the digital unitary evolution can be designed for the different Hamiltonians chosen in this work. For instance, one single step of TS decomposition for Eq.~\eqref{Ham_F} will look like,
\begin{equation}
\hat{U}(0,\Delta t) = e^{-i \hat{H}(t) \Delta t} \approx  e^{-i\theta_x(\Delta t)\sigma_x \Delta t} e^{-i\theta_z(\Delta t)\sigma_z \Delta t}e^{-i\theta_y(\Delta t)\sigma_y \Delta t},
\label{trott2}
\end{equation}
where $\theta_x(\Delta t) = (1- \lambda(\Delta t))$, $\theta_z(\Delta t) = \lambda(\Delta t)$ and $\theta_y(\Delta t) = F^{(1)}(\Delta t)$, are the variables that represent the change in the Hamiltonian in each step. Similarly, for Eq.~\eqref{Ham2qTot}, four variables are required for each spin in each step, i.e.,
\begin{gather}
\hat{U}(0,\Delta t) \approx  \prod_{j=1}^2 e^{-i \theta_x(\Delta t)\sigma_{x}^j \Delta t} e^{-i\theta_z(\Delta t)\sigma_{z}^j \Delta t}
e^{-i\theta_{zz}(\Delta t)\sigma_{z}^j \sigma_{z}^{j+1} \Delta t} e^{-i\theta_y(\Delta t)\sigma_{y}^j }.
\label{trot_final}
\end{gather}
These unitary operators are implemented in the circuit model. According to the Solovay-Kitaev theorem \cite{dawson2005solovay}, any k-body unitary operation can be decomposed into a combination of single-qubit and two-qubit gate operations. Following are some example of the implementations, corresponding to the unitary operators used in this study,

\begin{figure}[h]
    \centering
    \includegraphics[width=\linewidth]{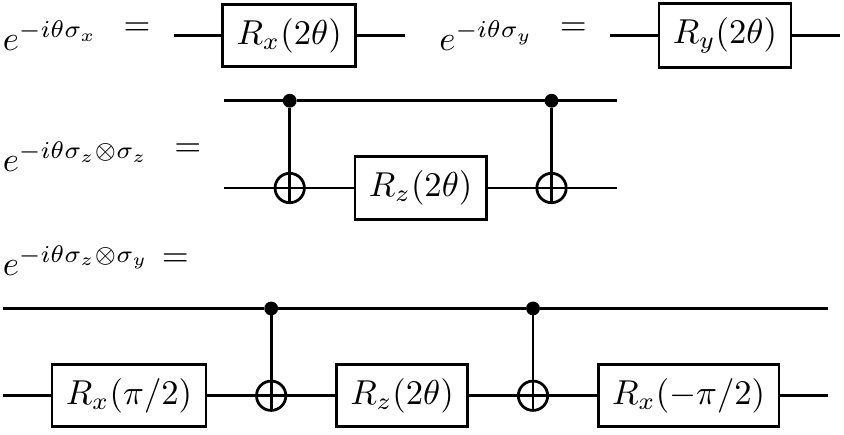}
\end{figure}




\section{Approximate CD term using variational method}\label{app2}
The main idea of counter-diabatic driving is to add an auxiliary term to the original Hamiltonian and evolve the system according to an effective Hamiltonian, 
\begin{equation}
    \hat{H}(t) = \hat{H}_0(t) + \dot{\lambda} \hat{A}_{\lambda},
\end{equation}
where $\hat{H}_0$ is the original Hamiltonian, $\dot{\lambda}$ is the control parameter, and $\hat{A}_{\lambda}$ is the exact adiabatic gauge potential responsible for the diabatic transitions. For the spin model considered in our simulation, the calculation of exact gauge potential results in non-local m-body interaction terms. Even though it is possible to implement these interactions using a basic set of quantum gates, the required gates will be huge and increase rapidly with the system size. Instead, for the practical purpose, we consider approximate gauge potential $\hat{A}_{\lambda}^*$, which satisfies the equation,
\begin{equation}
\left[i \partial_{\lambda} \hat{H}_0-\left[{\hat{A}}_{\lambda}^*, \hat{H}_0\right], \hat{H}_0\right]=0.
\end{equation}
For the optimal solution, we have to minimize the operator distance between the exact gauge potential and the approximate gauge potential, which is equivalent to minimizing the action,
\begin{equation}
{S_{\lambda}}\left({\hat{A}}_{\lambda}^{*}\right)=\operatorname{Tr}\left[\hat{G}_{\lambda}^{2}\left({\hat{A}}_{\lambda}^{*}\right)\right],
\end{equation}
where the Hilbert-Schmidt norm $\hat{G}_{\lambda}$ is given by,
\begin{equation}
\hat{G}_{\lambda}\left({\hat{A}}_{\lambda}^{*}\right)=\partial_{\lambda} \hat{H}_{0}+i\left[{\hat{A}}_{\lambda}^{*}, \hat{H}_{0}\right].
\end{equation}
A simple ansatz for $\hat{A}_{\lambda}^*$ for the Hamiltonian in Eq.~\eqref{non-int} from the main-text is, 
$\hat{A}_{\lambda}^{*}=\sum_{j} \alpha_{j}(t) \sigma_{y}^{j}$. 
This single qubit approximation works very well even for many-body systems. However, when the spin interaction terms become the leading term of the adiabatic gauge potential, this ansatz fails. So, we consider a general way to choose the ansatz using sequence of nested commutators proposed in \cite{sels2019prl}, that is,
\begin{equation}
\hat{A}_{\lambda}^{(l)} = i \sum_{k = 1}^l \alpha_k (t) \underbrace{[\hat{H},[\hat{H},......[\hat{H},}_{2k-1}\partial_{\lambda} \hat{H}]]], \label{gauge2}
\end{equation}
from which, when $l\rightarrow \infty$, we will get the exact gauge potential.

\begin{table}[]
\caption{The state fidelity using circuit optimization is depicted.}
\resizebox{\linewidth}{!}{%
\begin{tabular}{cccccc}
\hline
\hline
\multirow{2}{*}{\begin{tabular}[c]{@{}c@{}}Circuit \\ optimization\end{tabular}} & \multicolumn{2}{c}{Fidelity} & \multicolumn{2}{c}{Gate Count} & \multirow{2}{*}{\begin{tabular}[c]{@{}c@{}}Expected \\ gate error\end{tabular}} \\ \cline{2-5}
 & \multicolumn{1}{l}{Ideal} & \multicolumn{1}{l}{Experiment} & \multicolumn{1}{l}{Rotation} & \multicolumn{1}{l}{CNOT} &  \\ \hline
\multicolumn{6}{c}{Bell state preparation} \\ \hline
Optimized & \multirow{2}{*}{0.999} & 0.9835 & 8 & 2 & 0.01927 \\
Not optimized &  & 0.8021 & 19 & 14 & 0.11834 \\ \hline
\multicolumn{6}{c}{GHZ state preparation (3 qubit)} \\ \hline
Optimized & \multirow{2}{*}{0.9325} & 0.8198 & 20 & 7 & 0.07063 \\
Not optimized &  & 0.7370 & 23 & 15 & 0.14276 \\
\hline
\hline
\end{tabular}
}
\label{tab2}
\end{table}
\section{Error Analysis}\label{app3}
Various errors significantly impact the outcomes of the experiments. The sources of errors can be divided into three main categories. (i) {\it Discretization error} arises due to the choice of $\Delta t$ in the trotterization process, (ii) {\it Cumulative gate error} is a combination of single qubit gate errors and CNOT errors and increases linearly with the circuit depth and (iii) {\it Measurement error}  arises due to the measurements at the end of the time evolution. 
Also if the system evolves for a long time, as in the adiabatic case, the energy relaxation and dephasing also has to be considered. The cross-talks between the qubits and other environmental effects can also disturb our simulation, but these effects have not been considered in our simulations. 
\paragraph{Discretization error:}\label{app3_1}
In digital quantum simulation the main source of error arises from the discretization of the continuous time evolution of a Hamiltonian and decomposing this evolution into a sequence of quantum gates. The discretization can be performed using various methods, but the Trotter-Suzuki (TS) formula is the most widely used method among all because of its simplicity. In our simulation, we consider first order TS formula, where the error is of the order $\mathcal{O}(\Delta t^2)$. For a given total time $T$, we can choose an arbitrarily small value for $\Delta t$ to decrease the trotter error. However, with small $\Delta t$, we need more trotter steps to reach the final time, which will increase the total gate count and eventually leads to accumulation of gate error. One possible solution for this problem is to consider a higher-order TS formula using extra gates \cite{hatano2005finding}. Since the gate error is comparatively larger than the trotter error, we restrict ourselves to a first-order approximation.
\begin{table}[]
\caption{ We compare the number of quantum gates required for the successful implementation of adiabatic evolution on a digital quantum computer by including the CD term and excluding the CD term.}
\resizebox{\linewidth}{!}{%
\begin{tabular}{ccccc}
\hline
\hline
System & Trotter step & Rotation gates & CNOT gates & Fidelity \\ \hline
 & \multicolumn{4}{c}{With CD} \\ \hline
Single spin system & 2 & 7 & - & 0.995 \\
\begin{tabular}[c]{@{}c@{}}Non-integrable Ising model\\ (5-qubit, $J_0=-0.1$)\end{tabular} & 4 & 70 & 40 & 0.993 \\
Bell-state preparation & 3 & 27 & 14 & 0.999 \\
GHZ state (3 qubit) & 4 & 111 & 60 & 0.966 \\ \hline
\multicolumn{1}{c}{} & \multicolumn{4}{c}{Without CD} \\ \hline
Single spin system & 20 & 39 & - & 0.996 \\
\begin{tabular}[c]{@{}c@{}}Non-integrable ising model\\ (5-qubit, $J_0=-0.1$)\end{tabular} & 30 & 445 & 300 & 0.985 \\
Bell-state preparation & 24 & 70 & 48 & 0.998 \\
GHZ state (3 qubit) & 18 & 105 & 108 & 0.962 \\ \hline
\hline
\end{tabular}
}
\label{table3}
\end{table}
\paragraph{Gate error:}\label{app3_2}
While implementing the time evolution of a system, gate error plays a crucial role. With increasing trotter steps, the gate error also increases linearly. The average fidelity of a single qubit and a CNOT gate of IBM quantum computer in our simulation is $99.95\%$ and $98.5\%$, respectively. For the experimental implementation on a noisy device, it's necessary to optimize the quantum circuit before sending it to the hardware to get the desired result. In our simulation, to decrease the gate error, we used transpilation function available in Qiskit Terra for circuit optimization. Table~\ref{tab2} shows the experimental fidelity for the preparation of Bell-state and GHZ state with circuit optimization and without circuit optimization. The gate count and the expected gate error is calculated in both cases. In Table~\ref{table3}, the gate counts for the successful implementation of the adiabatic evolution for different systems on a digital quantum computer is illustrated. From the data, it is conclusive that the inclusion of the CD term can improve fidelity and reduce the total gate count.
\paragraph{Measurement error mitigation:}\label{app3_3}
One of the main source of error in our simulation is the readout error of the device. In the following, we briefly discuss how to mitigate measurement error on a small system using matrix inversion method. For that, we have to find out the response matrix $M_R$ for the given device. To measure $M_R$, we consider a set of $2^n$ calibration circuits using only X-gate. 
Let $P_{noisy}$ be the probability distribution for each possible $2^n$ states obtained from the quantum processor after measuring at the end. $P_{actual}$ be the probability distribution without readout noise. Then we can obtain $P_{mitigated}$ which is approximately equal to the $P_{actual}$ by applying the matrix inversion $M_R^{-1}$ on the obtained result, i.e.       
\begin{equation} 
    M_R^{-1} P_{noisy} = P_{mitigated}.
\end{equation}
This method works only when the measurement error is much larger than the single qubit gate error used for initial state preparation. This is true for IBMQ devices, where the average single qubit gate error is of the order of $10^{-4}$ and the measurement error is of the order $10^{-2}$. In this work we use the tool provided by qiskit ignis \cite{qiskit} for performing the measurement error mitigation. More advanced methods can be found in \cite{maciejewski2020mitigation}. 
\end{appendices}
\bibliography{main}

\end{document}